\documentclass[twocolumn,showpacs,preprintnumbers,amsmath,amssymb]{revtex4}
\usepackage{epsf}
\begin{document}

\title{Rectifying fluctuations in an optical lattice}
 
\author{P.H. Jones, M. Goonasekera and F. Renzoni}
 
\affiliation{Departement of Physics and Astronomy, University College London, 
Gower Street, London WC1E 6BT, United Kingdom}
 
\date{\today}
 
\begin{abstract}
We have realized a Brownian motor by using cold atoms in a dissipative optical 
lattice as a model system. In our experiment the optical potential is spatially
symmetric and the time-symmetry of the system is broken by applying 
appropriate zero-mean ac forces. We identify a regime of rectification
of forces and a regime of rectification of fluctuations, the latter 
corresponding to the realization of a Brownian motor.
\end{abstract}

\maketitle

Noise is unavoidably present in every physical, chemical and biological 
process. From electronic devices to laser action, from chemical reactions
to the motion of bacteria, noise modifies the process by introducing random 
fluctuations in the observed dynamics.

Noise is often treated as a nuisance to be reduced as much as possible. 
Consider for example an electronic device, typically
an amplifier, with an input and an output. It is often believed that the 
only way to improve the signal-to-noise ratio (SNR) at the output of the
device is to reduce the noise at its input. This is not necessarily true: for 
a nonlinear device the provision of additional noise to the input signal may
increase the SNR of the output signal, a phenomenon known as stochastic 
resonance and observed in a variety of processes in electronics, physics,
chemistry and biology \cite{stoch}.

Noise also plays a central role in Brownian motors \cite{ratchet}, the topic
of the present work, which recently attracted much interest
\cite{magnasco,flach,bartu,blanter,super,linke,harada}
as it is believed that they may constitute a model for biological molecular
motors \cite{prost}.  Consider a sample of Brownian particles diffusing 
through a periodic potential in the presence of oscillating forces of zero 
average. A net current of particles can be induced by breaking the symmetry 
of the system, realizing in this way the somewhat surprising situation of 
directed motion in a macroscopically flat potential in the absence of applied
bias forces. Whenever the net current of particles arises from the 
{\it rectification of fluctuations} the directed motion corresponds to the 
realisation of a so-called Brownian motor.

In this work we realize a Brownian motor by using cold atoms in a
dissipative optical lattice as a model system. The optical potential is 
spatially symmetric, and the time-symmetry of the system is broken by applying
appropriate zero-mean ac forces. We identify a regime of rectification of the
fluctuations, corresponding to the realization of a Brownian motor, and a 
regime of rectification of the forces.  In the regime of rectification of
the fluctuations, the current amplitude vanishes in the absence of noise, and 
shows a stochastic resonance-like behavior at increasing noise amplitude, 
therefore confirming that our device acts as a fluctuations rectifier. In 
the regime of rectification of the forces, the directed motion of the atoms
through the lattice is only due to deterministic forces. In this regime
the noise acts as a nuisance and correspondingly the current decreases for 
increasing amplitude of the noise.
\begin{figure}[ht]
\begin{center}
\mbox{\epsfxsize 3.in \epsfbox{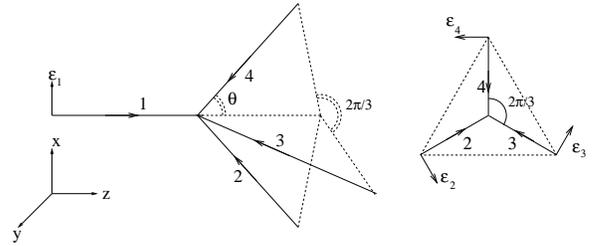}}
\end{center}
\caption{Lattice beams configuration. Left: arrangement of the four beams
in the umbrellalike configuration. Only the polarization of beam 1 is shown.
Right: polarizations of beams 2--4 as seen by looking towards the $-z$
direction.}
\label{fig1}
\end{figure}

The periodic potential used in this work is a 3D
optical lattice \cite{robi} determined by the 
interference of four laser beams (beams 1--4, with wavevectors 
$\vec{k}_1$--$\vec{k}_4$), arranged in the so-called umbrellalike
configuration \cite{petsas}, as sketched in Fig. \ref{fig1}. One laser beam
(beam 1) propagates in the $z$-direction. The three other laser beams 
propagate in the opposite direction, and are arranged along the edges of a 
triangular pyramid having the $z$-direction as axis, with the azimuthal angle 
between each pair of beams equal to $2\pi/3$. All beams are linearly 
polarized, with the polarization of beam 1 in the $x$-direction. The linear
polarization of beam $j$ is chosen as 
$\vec{\epsilon}_j=\vec{k}_1 \wedge \vec{k}_j/k^2$, i.e. 
$\vec{\epsilon}_j$ is orthogonal to the plane defined by ${\vec k}_1$ and 
${\vec k}_j$ and oriented as shown in Fig. \ref{fig1} (right). 
The angle between the beam $j$, with $j=$2--4, and the $z$ axis is 
$\theta=30^{\circ}$, and we have made the following choice for the 
fields amplitudes: $E_1={\cal E}_0$, $E_2=E_3={\cal E}_0
\sqrt{3+\cos^2\theta}/6\cos\theta={\cal E}_0\sqrt{5}/6$ and $E_4={\cal E}_0/3$.
The interference of the laser fields
produces a periodic and spatially symmetric optical potential, with the 
potential minima associated with pure circular ($\sigma^{+}$ or $\sigma^{-}$)
polarization of the light \cite{petsas}. For an atom with a $F_g=F\to 
F_e=F+1$ transition, the optical potential consists precisely of $2F+1$
potentials, one for each ground state sublevel of the atom. Transitions 
between different potentials are produced by optical pumping processes, 
which transfer an atom from one ground state sublevel to another one.
As optical pumping is a stochastic process, this introduces fluctuations
in the atomic dynamics. These fluctuations result in a random walk through
the optical lattice, and indeed normal diffusion has been observed for the 
atomic cloud expanding in the lattice for a broad range of interaction
parameters \cite{robi}.

In a spatially symmetric structure directed motion can be induced by 
breaking the time-symmetry of the system \cite{flach}. 
With this aim in mind, we apply a zero-mean ac
force composed of two harmonics
\begin{equation}
F(t)=F_0 \left[ A\cos (\omega t) + B \cos(2\omega t-\phi) \right]~.
\label{force}
\end{equation}
For $\phi\neq n\pi$, with $n$ integer, the force $F(t)$ breaks the 
time-symmetry of the system. To be precise, for an arbitrary choice of the 
phase $\phi$ the force $F(t)$ breaks the generalized-parity symmetry 
$(x,p,t)\to (-x,-p,t+\pi/\omega)$, and for $\phi\neq n\pi$ the residual 
time-reversal symmetry $(x,t)\to (x,-p,-t)$ is also broken \cite{flach}.
Experimentally, to introduce a homogenous time-dependent force we apply a 
phase modulation to the beam 1 of the form 
\begin{equation}
\alpha(t) = \alpha_0\left[ A\cos(\omega t)+\frac{B}{4} \cos(2\omega t-\phi)
\right]~.
\label{phase}
\end{equation}
In the accelerated frame in which the optical lattice is stationary the 
phase modulation results in an inertial force $F$ of the form of Eq.
\ref{force} with $F_0=m\omega^2\alpha_0/2k$, where $m$ is the atomic mass
\cite{schiavoni}. In the present work,
the phase difference between the two harmonics is kept fixed at $\phi=\pi/2$, 
so as to break the time-symmetry of the system and induce directed motion.
The coefficients $A$ and $B$ of the two harmonics will be taken 
as equal: $A=B=1$.

\begin{figure}[ht]
\begin{center}
\mbox{\epsfxsize 3.in \epsfbox{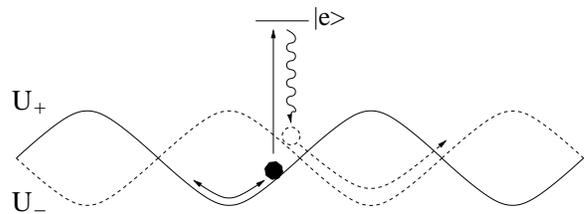}}
\end{center}
\caption{Optical potentials U$_{\pm}$ for a $J_g=1/2\to J_e=3/2$
atom in a 1D lin$\perp$lin optical lattice. A (stochastic) process of optical
pumping transferring, via an excited state, an atom from a potential to
the other one is also shown. The filled (empty) circle represents the atom in
the $|g,+\rangle$ ($|g,-\rangle$) state.}
\label{fig2}
\end{figure}

Our system differs from the usual models for
Brownian motors mainly in one respect: here we do not have a 
single potential, but a potential for each ground state of the atom, and both 
fluctuations in the atomic dynamics and the friction are associated with
optical pumping between ground states, i.e. optical pumping between
different optical potentials \cite{robi}. Because of this difference, we 
performed a numerical analysis prior to the experimental work. For simplicity,
our numerical analysis is limited to a 1D
lin$\perp$lin optical lattice \cite{robi} and a $J_g=1/2\to J_e=3/2$ atom.
This lattice beam geometry is the 1D version of our current experimental
setup, and corresponds to two counterpropagating laser beams with orthogonal
linear polarizations. In this case there are only two ground state sublevels
$|g,\pm\rangle$, which correspond to two sinusoidal optical potentials 
$U_{\pm}(z)$ in phase opposition. These potentials are shown in Fig.
\ref{fig2}, together with a (stochastic) optical pumping process which 
transfers an atom from one potential to the other inducing fluctuations in the
atomic dynamics. The Fokker-Planck-type equation describing the time evolution
of the semiclassical phase-space distribution has been derived in 
Ref. \cite{petsas2}. As also shown there, this equation can be efficiently
integrated by using Monte Carlo simulations techniques.

\begin{figure}[h]
\begin{center}
\mbox{\epsfxsize 3.in \epsfbox{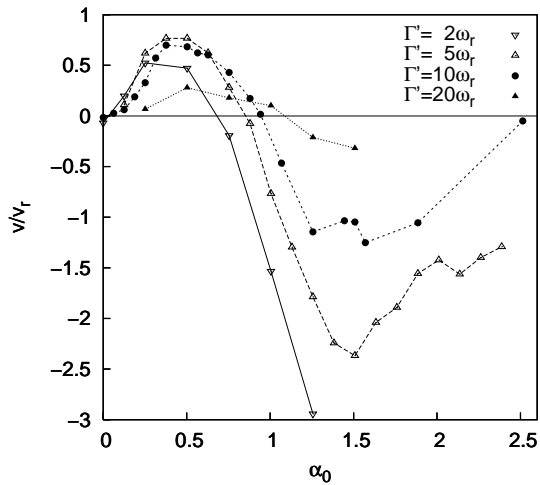}}
\end{center}
\caption{Results of semiclassical Monte Carlo simulations for a sample
of $n=10^4$ atoms in a 1D lin$\perp$lin optical lattice. The average
atomic velocity, in units of the recoil velocity $v_r =\hbar k/m$,
is shown as a function of the amplitude of the phase modulation.
Different data sets correspond to different optical pumping rates $\Gamma$'.
The lines are guides for the eyes.  The parameters of the calculations are:
the depth of the optical potential is $U_0=200/3\cdot E_r$, with $E_r$ the
recoil energy; the coefficients of the harmonics composing the phase
modulation, see Eq.~(\protect\ref{phase}), are equal: $A=B=1$, the
relative phase between the two harmonics is $\phi=\pi/2$, and the frequency
of the modulation is $\omega=0.92\cdot\omega_v$, where $\omega_v$ is the
vibrational frequency of the atoms at the bottom of the well.}
\label{fig3}
\end{figure}

In our calculations a phase modulation of one of the lattice beams of 
the form of Eq. (\ref{phase}) with $\phi=\pi/2$ is included to generate the 
appropriate ac force. We made Monte Carlo simulations for the atomic dynamics
and derived the mean atomic velocity as a function of the amplitude 
of the phase modulation, for a given optical potential depth and modulation 
frequency, at various optical pumping rates. From the numerical results,
shown in Fig. \ref{fig3}, it appears that a current is generated, a result 
which stimulates our experimental work. The dependence of the current amplitude 
on the optical pumping rate evidenced in the numerical simulations will be 
discussed once the experimental findings are presented. It should be noted 
that in reporting the numerical results (and the same will apply to the 
experimental findings) we do not distinguish between velocities in the 
laboratory frame, defined by the $z$ coordinate, and in the accelerated
frame, defined by $z'=z-\alpha(t)/(2k)$, in which the optical 
potential is stationary. This because the two velocities coincide once 
averaged over time scales $T$ much larger than the ac forces period.
The typical frequency $\omega$ considered in this work is about 
100 kHz, while the typical averaging time is larger than 1 ms, so the 
average velocities in the laboratory and accelerated frames are equal.

In our experiment cesium atoms are cooled and trapped in a magneto-optical
trap. At a given instant the trap is switched
off and the four lattice beams are turned on. The lattice fields are red 
detuned with respect to the $F_g=4\to F_e=5$ D$_2$ line. The phase modulation 
$\alpha(t)$, see Eq.~(\ref{phase}), of beam 1 is then slowly turned on. 
The modulating signal is obtained by adding the output signals of two 
phase-locked oscillators, with oscillation frequencies $\omega$ and $2\omega$
and phase difference $\phi=\pi/2$. We observed the motion of the atoms in the 
lattice by direct imaging of the atomic cloud with a CCD camera. Consistenly
with previous work \cite{flach,schiavoni,chaos} we observed directed motion of
the atoms through the lattice along the $z$ direction, following the 
time-symmetry breaking. We observed a uniform motion of the center of mass
of the atomic cloud, and we derived from the experimental data a mean atomic
velocity. Several sets of 
measurements were made for different choices of the lattice beams' parameters
(intensity $I$ and detuning $\Delta$ from atomic resonance), and modulation
amplitude $\alpha_0$. The lattice beams' intensity and detuning were varied 
simultaneously to keep constant the depth of the optical potential
$U_0\propto I/\Delta$, as verified by pump-probe spectroscopy, while changing
the optical pumping rate $\Gamma'\propto I/\Delta^2$. 
In this way we have measured, for a given optical
potential, the average atomic velocity as a function of the modulation
amplitude, i.e. as a function of the ac force amplitude, for different 
optical pumping rates, i.e. for different noise levels.

\begin{figure}[htb]
\begin{center}
\mbox{\epsfxsize 3.in \epsfbox{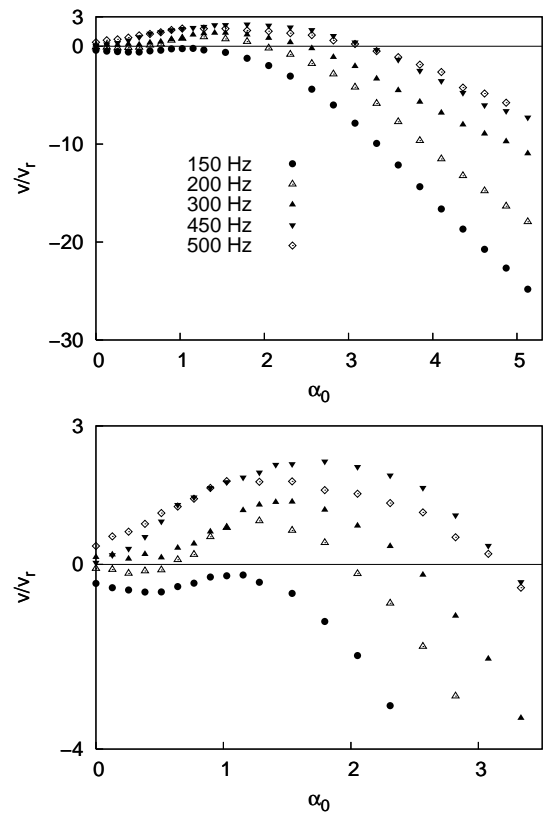}}
\end{center}
\caption{Experimental results for the mean atomic velocity
as a function of  the amplitude of the phase
modulation. The top graph includes all our experimental results, while the
bottom
one evidences the region of small ac forces.
The optical potential is the same for all measurements, and corresponds
to a vibrational frequency $\omega_v=2\pi\cdot 144$ kHz.
Different data sets correspond to different detunings $\Delta$, i.e. to
different optical pumping rates as the optical potential
is kept constant. The data are labelled
by the quantity $\Gamma_s = [\omega_v/(2\pi)]^2/\Delta$ which is proportional
to the optical pumping rate. The modulation frequency is
$\omega=2\pi\cdot 130$kHz. The values for the velocity are expressed in terms
of the recoil velocity $v_r$, equal to 3.52 mm/s for the the Cs D$_2$ line.
Each datapoint corresponds to an average over five images of
the atomic cloud. In plotting the data sets for the different pumping rates,
adjacent averaging on each data set has also been made. As unwanted
consequence,
a small vertical shift (positive or negative depending on the data set) for
the datapoints close to $\alpha_0=0$ has been introduced by the averaging
procedure. Errorbars are of the order of the size of the datapoints.}
\label{fig4}
\end{figure}

Results of our measurements are reported in Fig.  \ref{fig4}. The experimental
data show the same behavior as our numerical results (see Fig. \ref{fig3}).
For small amplitudes of the ac force the average atomic velocity is an 
increasing function of the force amplitude, with the atoms moving in the 
positive $z$ direction.  At larger amplitudes of the ac force the velocity
decreases, and a current reversal is observed, with the atomic cloud moving in
the negative $z$ direction. We note that the numerical simulations (see
Fig.~\ref{fig3}) show that at large ac forces amplitude the current reaches
a maximum and
then decreases at increasing amplitude of the ac forces. This has a simple 
explanation \cite{bartu}: for very large amplitudes of the applied ac forces
the influence of the periodic potential on the atomic dynamics becomes
negligible, and the current of atoms decreases. This behavior has not been
observed in the experiment because, due to technical limitations, we were not
able to explore modulation amplitudes $\alpha_0$ large enough. 

It is important to distinguish the different mechanisms leading to current 
generation. For large amplitudes of the applied force, the motion can be 
attributed to deterministic forces and corresponds to force rectification
by harmonic mixing: in a nonlinear medium the two harmonics, of frequency
$\omega$ and $2\omega$ and phase difference $\phi$, are mixed and result
in a {\it rectified force}  $\bar{F}\propto \sin\phi$. In our experiment
the nonlinearity of the medium is the nonlinearity of the optical potential.
In this regime the noise does not play any constructive role in the generation
of the current of atoms. On the contrary, the noise disturbs the process of
rectification of the forces, and, as it appears from our experimental data,
for large ac forces the average atomic velocity decreases for increasing
optical pumping rate, i.e. for increasing level of the noise, in agreement
with our numerical simulations (Fig. \ref{fig3}). We therefore conclude that 
for large applied forces our experimental realization does not correspond to 
a Brownian motor.

Consider now the case of small
applied ac forces. As evidenced by our experimental data (Fig.~\ref{fig4},
bottom) the dependence of the average atomic velocity  
on the optical pumping rate is completely different from the one 
observed at large ac forces. For small values of the optical
pumping rate the current amplitude is an increasing function of the pumping 
rate, and the current vanishes in the limit of vanishing optical
pumping rate, as shown by the filled circles datapoints of Fig.~\ref{fig4}
(bottom) which correspond to the smallest value of the optical pumping rate
explored in our experiment. Finally, at larger pumping rates the current 
reaches a maximum and then decreases. This stochastic resonance-like 
behavior is the demonstration that in the reverse current regime our 
optical lattice acts as a fluctuations rectifier, i.e. we realized a 
Brownian motor.

In conclusion, in this work we presented the realization of a Brownian motor 
by using cold atoms in a dissipative optical lattice as a model system. We 
considered a spatially symmetric optical lattice, and we broke the 
time-symmetry of the system by applying an appropriate zero mean ac force.
A current of atoms is generated as a result of the 
time-symmetry breaking. We identify two different regimes, depending on the 
amplitude of the applied ac force. For large amplitudes the current
is produced by deterministic forces, and can be traced back to the 
rectification by harmonic mixing of the applied oscillating forces. In this 
regime the noise acts as a disturbance for the rectification process, and 
correspondingly the current amplitude is a decreasing function of the noise
level. At small amplitude of the ac force, the current is reversed and is due
to the rectification of fluctuations, with the current amplitude showing a 
stochastic resonance-like dependence on the noise level. This corresponds
to the realization of a Brownian motor.

The present work also shows the important role that optical lattices can play
in generating models for statistical physics. With respect to solid state 
devices \cite{linke} or laser tweezer setups \cite{harada}, optical
lattices offer a wider tunability. The depth of the defect-free optical
potentials can be controlled by simply changing the laser parameters, and by
changing the arrangement and the number of laser beams both periodic
and quasi-periodic lattices of different dimension, lattice spacing and
lattice geometry can be obtained. Furthermore, the laser parameters also allow
a precise control of the optical pumping rate, which can be varied over a very
broad range, and also eventually be completed suppressed. This allows the
investigation of the vast field of noise-induced phenomena. Among the 
phenomena identified theoretically that can be explored by our current set-up
we mention: dissipation-induced symmetry breaking \cite{super}, giant 
acceleration of free diffusion in tilted lattices \cite{giant}, Levy 
walks and anomalous diffusion \cite{levy}. Furthermore, in the limit of 
far detuning from atomic resonance, i.e. by suppressing dissipation, 
deterministic (chaotic) ratchets \cite{jung} can be investigated.


\begin{thebibliography}{99}
\bibitem{stoch}
K.~Wiesenfeld and F.~Moss, Nature {\bf 373}, 33 (1995);
L.~Gammaitoni {\it et al.}, Rev. Mod. Phys. {\bf 70}, 223 (1998);
M.I.~Dykman and P.V.E.~McClintock, Nature {\bf 391}, 344 (1998).
\bibitem{ratchet}
R.D.~Astumian and P.~H\"anggi, Phys. Today {\bf 55}, 33 (2002);
P.~Reimann, Phys. Rep. {\bf 361}, 57 (2002);
P.~Reimann and P.~H\"anggi, Appl. Phys. A {\bf 75}, 169 (2002);
see also P.V.E.~McClintock, Nature {\bf 401}, 23 (1999).
\bibitem{magnasco}
M.O.~Magnasco, Phys. Rev. Lett. {\bf 71}, 1477 (1993).
\bibitem{flach}
A.~Adjari {\it et al.},
J. Phys. I (France) {\bf 4}, 1551 (1994);
M.C.~Mahato and A.M.~Jayannavar, Phys. Lett. A {\bf 209}, 21 (1995);
D.R.~Chialvo and M.M.~Millonnas, {\it ibid.} {\bf 209}, 26 (1995);
M.I.~Dykman {\it et al.}, Phys. Rev.  Lett. {\bf 79}, 1178 (1997);
S.~Flach, O.~Yevtushenko and Y.~Zolotaryuk, {\it ibid.} {\bf 84},
2358 (2000); S.~Denisov {\it et al.}, Phys. Rev. E {\bf 66}, 041104 (2002);
M.V.~Fistul, A.E.~Miroshnichenko and S.~Flach, Phys. Rev. B {\bf 68}, 153107
(2003); I.~Goychuk and P.~H\"anggi,  Europhys. Lett. {\bf 43}, 503 (1998).
\bibitem{bartu}
R.~Bartussek, P.~H\"anggi and J.G.~Kissner, Europhys. Lett. {\bf 28},
459 (1994).
\bibitem{blanter}
Ya.M.~Blanter and M.~B\"uttiker, Phys. Rev. Lett. {\bf 81}, 4040 (1998).
\bibitem{super}
P.~Reimann, Phys. Rev. Lett. {\bf 86}, 4992 (2001);
O.~Yevtushenko {\it et al.}, Europhys. Lett. {\bf 54}, 141 (2001).
\bibitem{linke}
H.~Linke {\it et al.}, Europhys. Lett. {\bf 44}, 341 (1998);
S.~Weiss {\it et al.}, {\it ibid.} {\bf 51}, 499 (2000).
\bibitem{harada}
T.~Harada and K. Yoshikawa, Phys. Rev. E {\bf 69}, 031113 (2004).
\bibitem{prost}
F.~J\"ulicher, A.~Ajdari and J.~Prost, Rev. Mod. Phys. {\bf 69}, 1269 (1997).
\bibitem{robi}
For a recent review of optical lattices, see
G.~Grynberg and C.~Mennerat-Robilliard, Phys. Rep. {\bf 355}, 335 (2001).
\bibitem{petsas}
K.I.~Petsas, A.B.~Coates, and G.~Grynberg, Phys. Rev. A {\bf 50}, 5173 (1994).
\bibitem{schiavoni}
M.~Schiavoni {\it et al.}, Phys. Rev.  Lett. {\bf 90}, 094101 (2003).
\bibitem{petsas2}
K.I.~Petsas, G.~Grynberg and J.-Y.~Courtois, Eur. Phys. J. D {\bf 6}, 29 (1999).
\bibitem{chaos}
M.I.~Dykman {\it et al.}, Chaos {\bf 11}, 587 (2001).
\bibitem{giant}
P.~Reimann {\it et al.}, Phys. Rev. Lett. {\bf 87}, 010602 (2001)
\bibitem{levy}
S.~Marksteiner, K.~Ellinger, and P.~Zoller, Phys. Rev. A {\bf 53}, 3409 (1996).
\bibitem{jung}
P.~Jung, J.G.~Kissner and P.~H\"anggi, Phys. Rev. Lett. {\bf 76}, 3436 (1996).
\end{thebibliography}
\end{document}